\documentclass[a4paper,11pt]{article}
\usepackage{pos}

\usepackage{graphicx}
\usepackage{multirow} 
\usepackage{amsmath}
\usepackage{enumerate}
\usepackage{subcaption} 
\usepackage{color} 
\usepackage[normalem]{ulem}
\usepackage[utf8]{inputenc}
\usepackage[outdir=./]{epstopdf}

\newcommand{\xmax}{\ensuremath{X_{\rm max}\,}}

\newcommand{\zhaires}{\mbox{ZHA{\scriptsize{${\textrm{IRE}}$}}{S }}}
\newcommand{\zhairesns}{\mbox{ZHA{\scriptsize{${\textrm{IRE}}$}}{S}}}

\title{RDSim, a fast and comprehensive simulation of radio detection of air showers}

\author*[a]{Washington  R. de Carvalho Jr.}
\author[a]{Abha Khakurdikar}

\affiliation[a]{IMAPP, Radboud University,\\ Heyendaalseweg 135, Nijmegen, The Netherlands\\ }

\emailAdd{carvajr@gmail.com}
\emailAdd{a.khakurdikar@astro.ru.nl}

\abstract{We present RDSim, a fast and comprehensive framework for the simulation of the radio emission and detection of downgoing air showers. It can handle any downgoing shower that can be simulated with \zhairesns, including those induced by CC and NC neutrino interactions and $\tau$ decays. RDSim is based on a superposition toymodel that disentangles the Askaryan and geomagnetic components of the shower emission.  By using full \zhaires simulations as input, it is able to estimate the full radio footprint on the ground. A single input simulation at a given energy and arrival direction can be scaled in energy and rotated in azimuth by taking into account all relevant effects. This makes it possible to simulate a huge number of geometries and energies using just a few \zhaires input simulations. The framework takes into account the main characteristics of the detector, such as trigger setups, thresholds and antenna patterns. To accommodate arrays that use particle detectors for triggering, such as the Auger RD extension, it also features a second toymodel to estimate the muon density at ground level, which is used to perform simple particle trigger simulations. It's speed makes it possible to investigate in detail events with a very low trigger probability, as well as many geometrical effects due to the array layout. In case more detailed studies of the radio detection are needed, RDSim can also be used to sweep the phase-space for the efficient creation of dedicated full simulation sets. This is particularly important in the case of neutrino events, that have extra variables that greatly impact shower characteristics, such as interaction or $\tau$ decay depth as well as the type of interaction and it's fluctuations.}

\FullConference{%
  9th International Workshop on Acoustic and Radio EeV Neutrino Detection Activities - ARENA2022\\
  7-10 June 2022\\
  Santiago de Compostela, Spain}

\begin{document}
\maketitle

\section{Introduction}



Initially proposed in the sixties, the radio detection of cosmic rays has undergone a renaissance in the last 15 years. It has now come of age as it has been shown to be competitive with other detection techniques while offering many advantages over them. It is currently being used by several cosmic ray and neutrino experiments worldwide~\cite{FrankRadioReview}.  

In this context we introduce RDSim, a framework for the simulation of the radio emission of extensive air showers (EAS) and its detection by an arbitrary antenna array. It is being developed with speed in mind and uses simple, yet still precise, toymodel-like approaches to simulate both the radio emission and the detector response. After an initial set up, it is able to simulate in detail millions of events in just a few minutes. This speed makes it possible to investigate larger areas around the detector, study events with very low detection probability and examine geometrical effects, such as border effects and those that arise due to asymmetries in the radio array. Thanks to the large statistics it makes possible, RDSim is specially suited to be used as a fast and accurate aperture calculator.

  
  

  This work is organized as follows: Section \ref{sec:emissionanddetection} describes the radio emission and detector response models used, including trigger parameters and a brief description of the optional particle trigger simulation.
Section \ref{sec:neutrinos} outlines the extra models used in the case of neutrino events, such as sampling of the neutrino interaction point and tau-lepton propagation. Section \ref{sec:structure} describes the general structure and procedures of RDSim, which is followed by a discussion on Section \ref{sec:discussion}.

\section{Radio emission and detector response modeling}
\label{sec:emissionanddetection}






  
\begin{figure}[!htb]
  \begin{center}
    \vspace{-0.6cm}
    \includegraphics[width=0.36\linewidth, trim= 0 -2cm  0 0.8cm, clip]{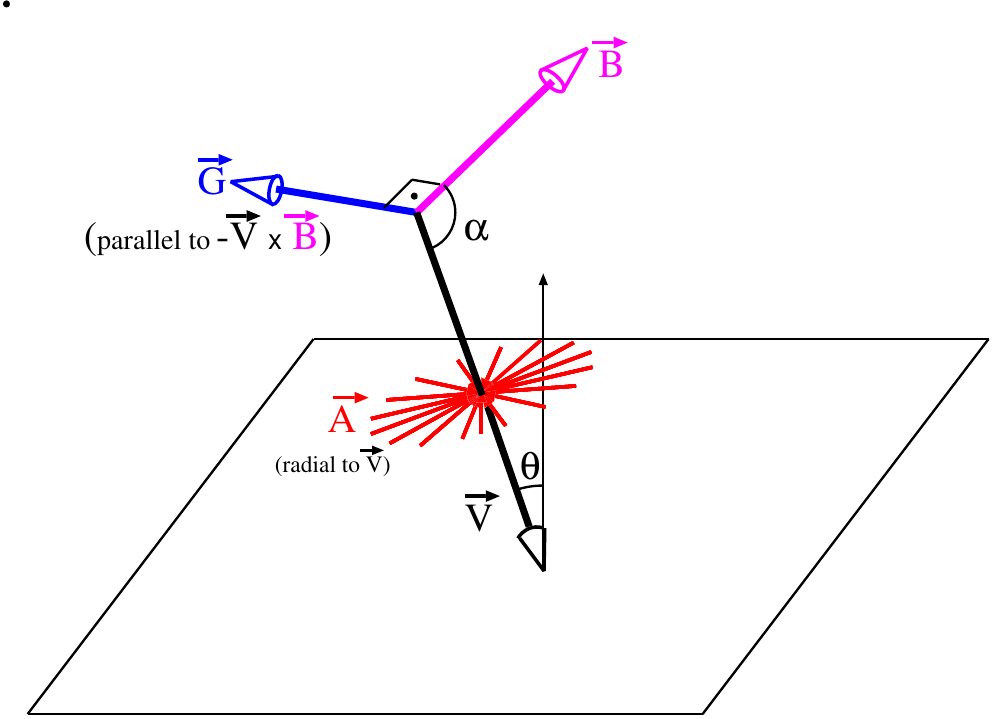}\hspace{1cm}\includegraphics[width=0.41\linewidth]{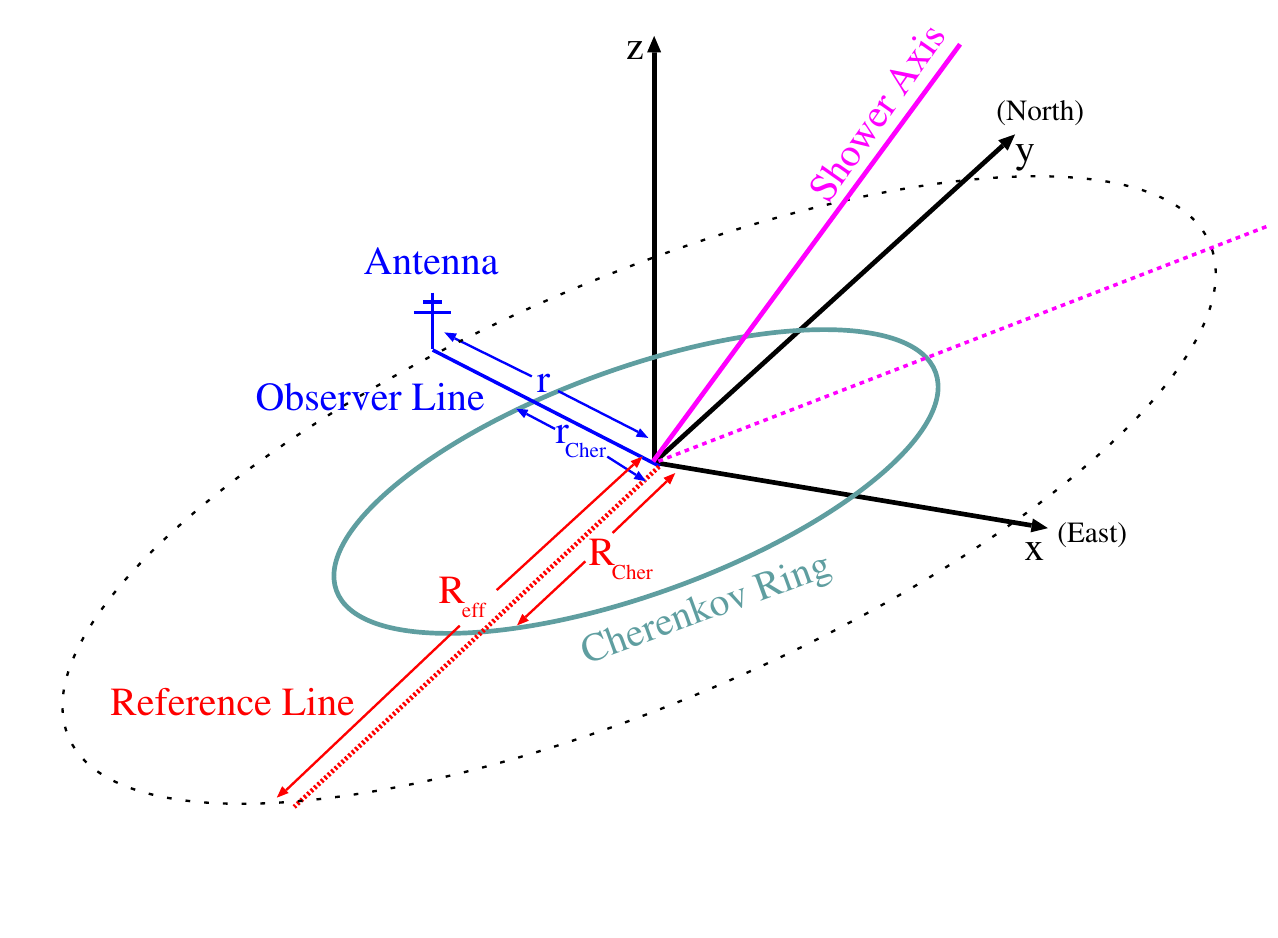}
  \end{center}
  \vspace{-0.6cm}
  \caption{Left: Theoretical polarization of the Askaryan (red) and geomagnetic (blue) emission mechanisms. The shower axis $\vec{v}$ is shown in black and the geomagnetic field $\vec{B}$ in magenta. Right: Assumed elliptical symmetry at ground level, described by an ellipse with its major axis aligned with the projection of the shower axis on the ground. The reference line, where the Askaryan and geomagnetic amplitudes are obtained from the input ZHAireS simulation is shown in red and the antenna where we want to estimate the electric field using the model is shown in blue. Both figures were extracted from \cite{toymodel}. }
  \label{fig:toymodelscheme}
\end{figure}

  The radio emission model in RDSim is based on the superposition of the Askaryan and geomagnetic emission mechanisms and is an expansion of the model presented in \cite{toymodel}. It uses as input full ZHAireS~\cite{zhaires} simulations of just a few antennas along a reference line. The superposition model then disentangles the Askaryan and geomagnetic components in order to get the amplitudes of the peak electric field, for each of the emission mechanisms separately, as a function of the distance to the core along the reference line. By assuming an elliptical symmetry for the amplitudes of each mechanism and using their theoretical polarizations (see Fig.~\ref{fig:toymodelscheme}), it is able to estimate the net peak electric field, along with its polarization, at any position on the ground. Given an arbitrary observer at a distance $r$ from the center of the ellipse (blue antenna on Fig.~\ref{fig:toymodelscheme}), we use the elliptical symmetry to get the distance $R_{\mbox{eff}}$ along the reference line (red line in Fig.~\ref{fig:toymodelscheme}) where we sample the Askaryan and geomagnetic amplitudes. We then add them up, taking into account their expected theoretical polarizations, to obtain the net electric field and polarization at the desired observer position (see \cite{toymodel} for more details).
  
\begin{figure}[!htb]
  \begin{center}
    \includegraphics[width=0.333\linewidth]{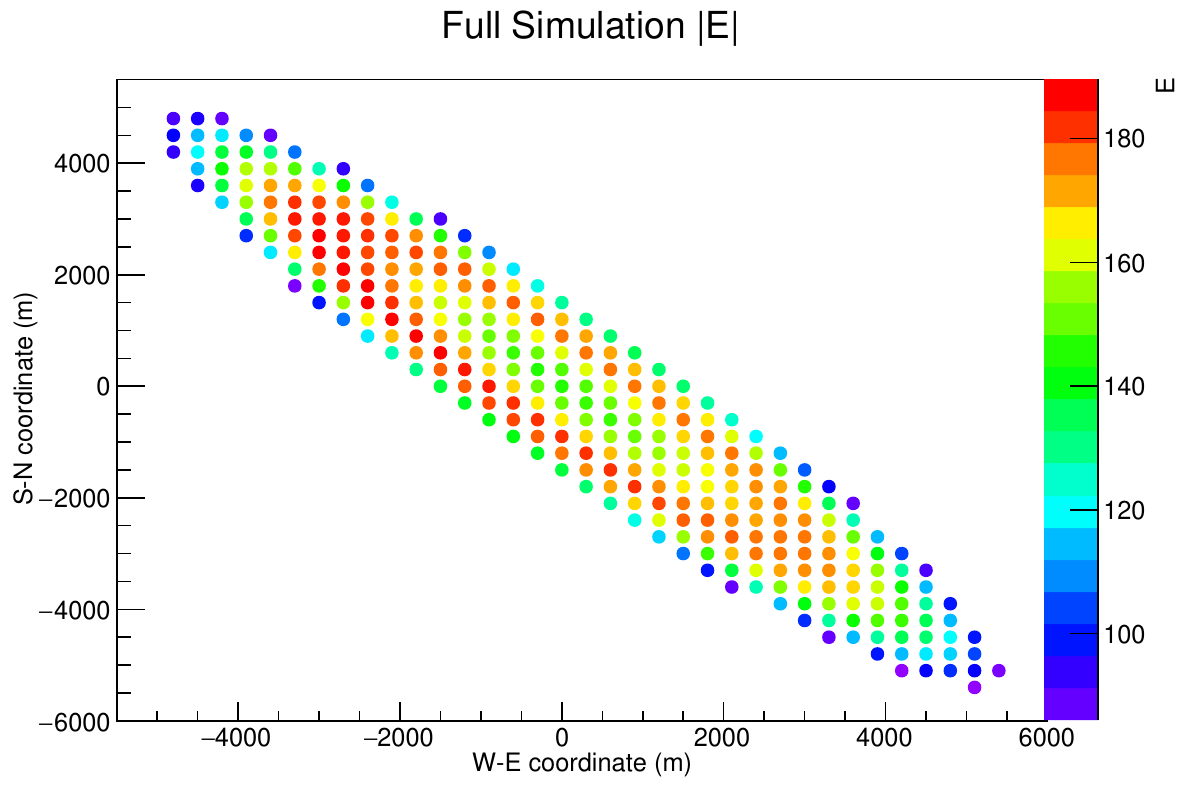}\includegraphics[width=0.333\linewidth]{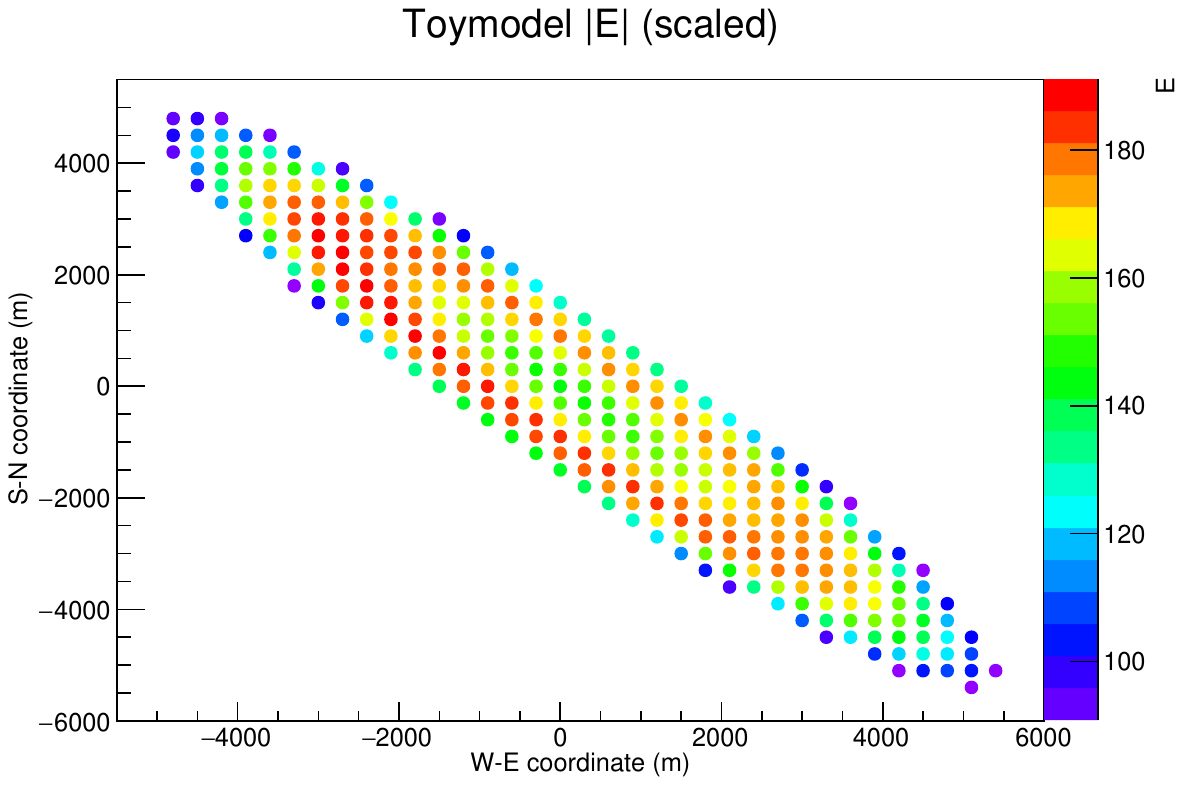}\includegraphics[width=0.333\linewidth]{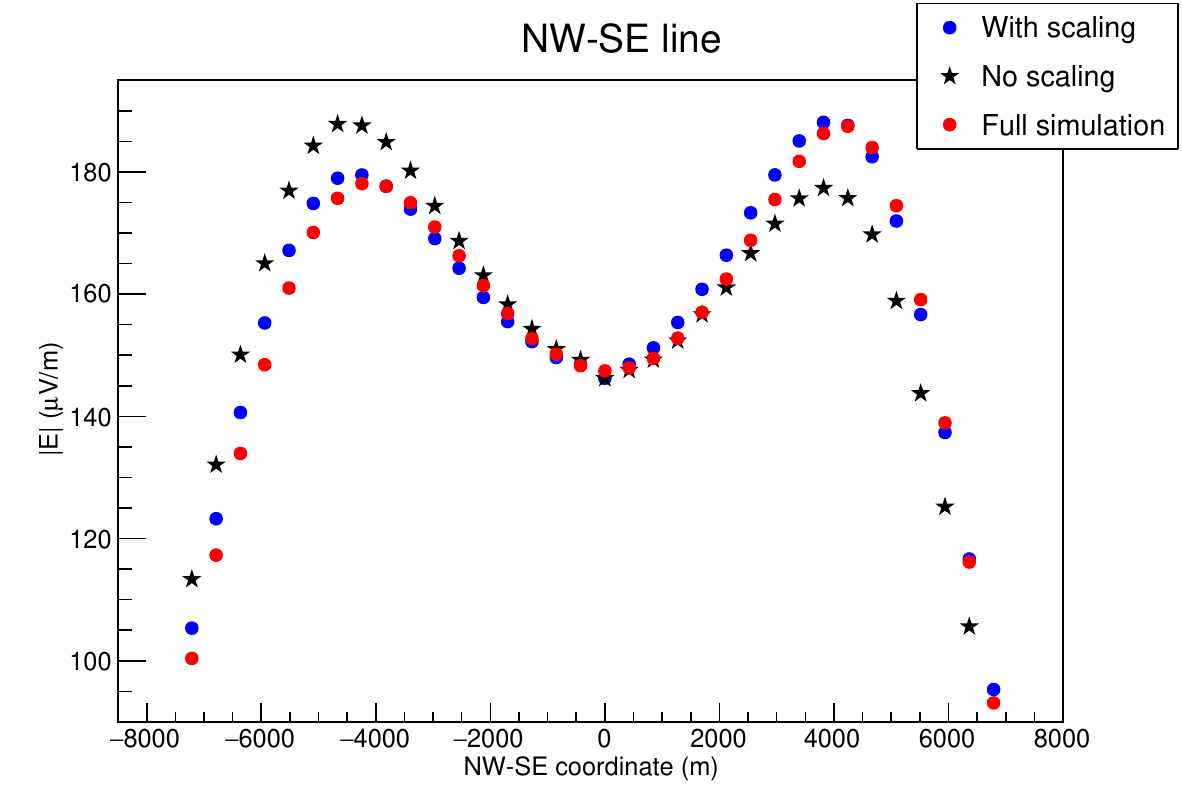}
  \end{center}
  \vspace{-0.6cm}
  \caption{Comparison between full simulations and the superposition model. See text for the details. The maximum difference in this example is 6\%.}
  \label{fig:fullsimcomparison}
\end{figure}

The original superposition model described in \cite{toymodel} did not take Early-Late effects into account and started to become inaccurate for showers with $\theta>70^\circ$ (see Fig. 7 of~\cite{toymodel} and the right panel of Fig.~\ref{fig:fullsimcomparison}, where the old model is labeled as ``No scaling''). In this new iteration of the model we now take into account the changes in the distance to the shower as the position of the observer changes. This is modeled by scaling both the Askaryan and geomagnetic amplitudes at the relevant point $R_{\mbox{eff}}$ on the reference line by $D^{\mbox{Ref}}_{\xmax}/D^{\mbox{Obs}}_{\xmax}$, where $D^{\mbox{Ref}}_{\xmax}$ is the distance between \xmax and $R_{\mbox{eff}}$ and $D^{\mbox{Obs}}_{\xmax}$ is the distance between \xmax and the observer.

We have compared the results of our expanded superposition model with full simulations of the radio emission. An example can be seen on Fig.~\ref{fig:fullsimcomparison}. The left panel shows the results of a full ZHAireS simulation of an $80^\circ$ shower and the middle panel shows the estimated field using the new model at the same positions. On the right panel we show the amplitudes of the electric field along the major axis of the elliptical radio footprint, where the Early-Late effects are maximum. One can see that the new model (marked as ``With Scaling'') has a very good agreement with the full simulation. The maximum difference in this example comparison is 6\%.


\begin{figure}[!htb]
  \begin{center}
   \includegraphics[width=0.48\linewidth,height=0.15\linewidth]{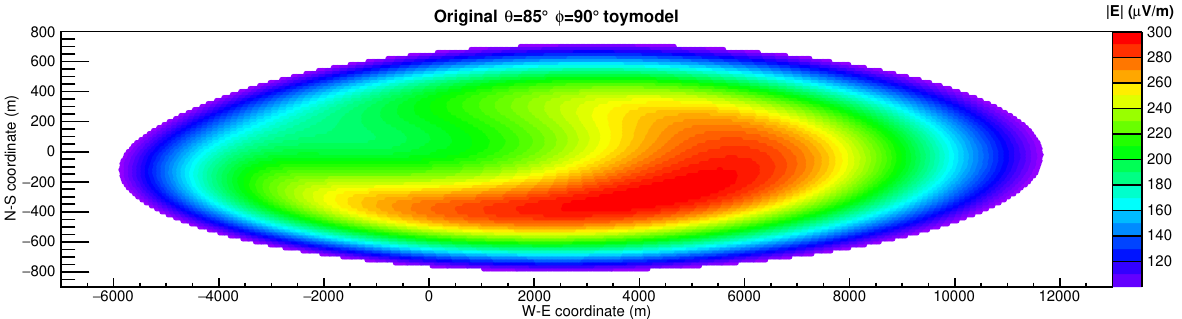}\includegraphics[width=0.52\linewidth,height=0.15\linewidth]{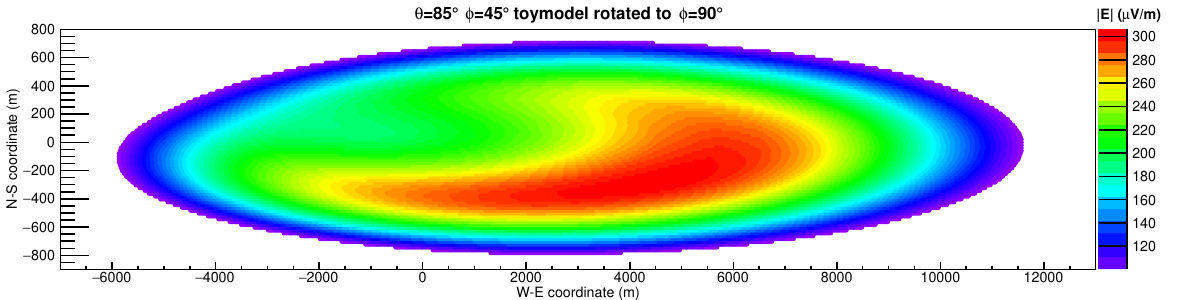}
  \end{center}
  \vspace{-0.5cm}
  \caption{Comparison between a dedicated toymodel with $\theta=85^\circ$ coming from the West (left) and a rotated toymodel, originally constructed as coming from the NW (right). The maximum difference in this example comparison is 2\%.}
  \label{fig:toyrotation}
\end{figure}

The old model could only estimate the net electric field for the exact same arrival direction as the ZHAireS simulation used to construct it. We have implemented a way to rotate the new superposition model to any desired azimuth angle, making it possible to reuse a single input ZHAireS simulation multiple times. In order to do this we rotate the ellipse to match the azimuth of the new desired arrival direction. But here we also have to take into account the changes in the angle $\alpha$ between the shower axis and the magnetic field (see left panel of Fig.~\ref{fig:toymodelscheme}), which has an impact on the amplitude of the geomagnetic component of the emission, as it roughly scales with $\sin{\alpha}$. To correct for this we scale the geomagnetic amplitudes along the whole reference line by $\sin{\alpha'}/\sin{\alpha}$, where $\alpha$ ($\alpha'$) is the angle between the original (rotated) shower axis and $\vec{B}$. We have found that the errors introduced by this rotation are very small, as can be seen in Fig.~\ref{fig:toyrotation}, where we compare the fields obtained by an unrotated model of a shower with $\theta=85^\circ$ coming from the West (left) with one rotated in azimuth by 45$^\circ$ to match that same arrival direction (right). The maximum difference in this example is just 2\%. We have also implemented a simple linear scaling of the electric field with shower energy, extending even further the phase-space that can be covered by a single ZHAireS input simulation.

The characteristics of the detector, including its response, is modeled in a very simple way. The antenna positions on a flat plane defined by the ground altitude at the array are arbitrary and read from an input text file. In the case of arrays that do not measure the vertical component of the electric field, RDSim can be set to just use its horizontal component. Antenna triggers are modeled by a simple settable threshold in electric field amplitude (default is 100 $\mu V/m$). The effect of the beam pattern of the antennas can also be taken into account by setting an optional input file. By default we assume the pattern is the same for each detectable polarization and we only take into account the zenithal dependence of the beam pattern w.r.t. the radiation arrival direction. For a given antenna, we then just multiply the original electric field obtained from the superposition model by the beam pattern, obtaining an ``effective'' electric field, which is then used instead of the original field for the simple threshold trigger. An array-level trigger is also implemented in a very simple way, by a settable minimum number of antenna-level triggers required to consider the whole event as triggered. At the end of the run, the information of all events is  saved to a compressed ROOT file, including the components of the measured electric field of each triggered antenna. This makes it possible to implement more complex analyses outside the simulation, e.g. more complex triggers, such as the ``veto antenna'' approach of the radio only-trigger at OVRO-LWA~\cite{ovroupdates}, or the simple particle trigger described below.

For detectors that require ground particles in order to trigger, we have implemented a simple particle trigger simulation. At the moment we only model high energy muons arriving at the ground. This particle trigger is based on a simple model to approximate the muon density at ground level, which uses low thinning AIRES~\cite{aires} simulations as input. As is the case with the radio emission model, we implemented a rotation of the muon model in order to use a single AIRES simulation to approximate the muon density for many arrival directions. More details can be seen in \cite{RDSim-ECRS}. The shape of the particle detector is taken into account by using an effective area $A_{\mbox{eff}}(\theta)$ as a function of the shower zenith angle, which is akin to a shadow area of the detector on the ground. For a circular water tank like the ones used in AUGER~\cite{AugerPrime},   $A_{\mbox{eff}}(\theta)=\pi r^2 + 2rh\tan{\theta}$, where $r$ is the tank radius and $h$ is the height of the water inside it. In the case of a horizontally installed scintillator at ground level, the effective area is just the geometrical area of the detector. We then estimate the number of muons crossing the detector by sampling a poissonian distribution with a rate parameter $\lambda=A_{\mbox{eff}}\rho_\mu$, where $\rho_\mu$ is the muon density at the location of the detector. The particle detector is considered triggered if the number of sampled muons crossing it is greater than a settable threshold. In order to maintain RDSim's great speed, this calculation is done after the main radio simulation is finished. This means that the particle trigger is calculated only for the events and stations that actually triggered in the radio-only part of the simulation.


\section{Extra modeling for Neutrino events}
\label{sec:neutrinos}

The emission model used in RDSim can handle any downgoing shower that can be simulated with ZHAireS. This includes neutrino CC and NC induced showers as well as those initiated by tau-lepton decays in the atmosphere. To simulate CC and NC initiated showers we use a previously produced extensive library of Herwig~\cite{Herwig} simulations of neutrino interactions and inject the secondaries into ZHAireS, which then simulates the shower and its radio emission. In the case of showers initiated by tau decays the procedure is the same, but instead of Herwig, we use our library of tau decays simulated with Tauola~\cite{Tauola}.

\begin{figure}
  \begin{center}
    \vspace{-0.5 cm}
    \includegraphics[width=0.5\linewidth]{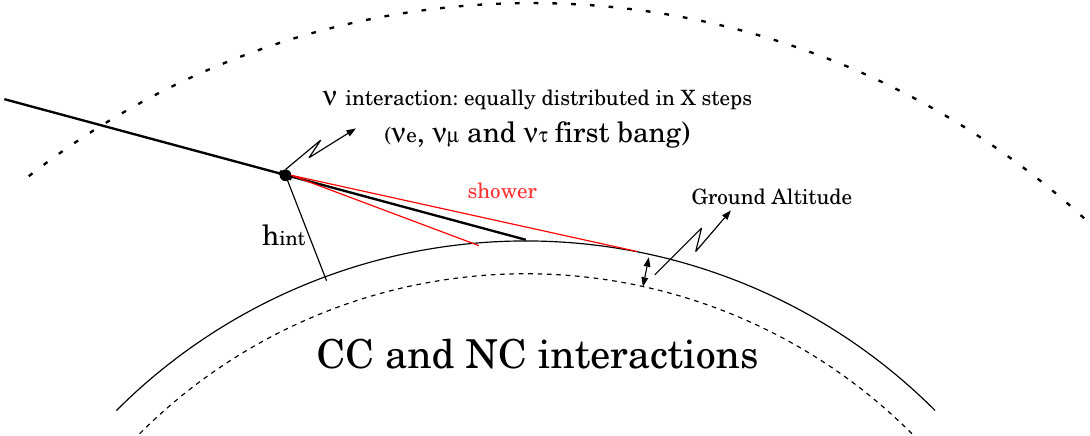}\includegraphics[width=0.5\linewidth]{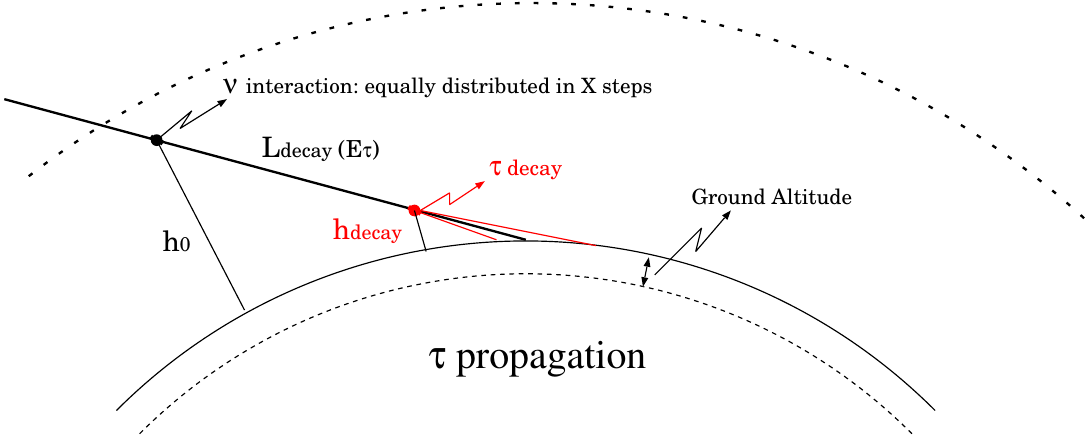}
  \end{center}
  \vspace{-0.6cm}
  \caption{Geometrical scheme of neutrino induced shower simulations. In both cases the atmosphere is divided in slices of equal thickness $\Delta X$ in atmospheric depth and the neutrino interaction is assumed to be equally probable in any of the slices. Left: CC and NC induced events. The neutrino interaction occurs at the center of a slice at a height $h_{\mbox{int}}$ above ground level and creates a shower. The height $h_{\mbox{int}}$, used internally by RDSim, is completely equivalent to the depth $X_{\mbox{int}}$ of the interaction. Right: Tau propagation. A $\nu_\tau$ interacts at a height $h_{0}$ above ground level (equivalent to a depth $X_{0}$). The tau is then propagated until it decays at a height $h_{\mbox{decay}}$ above ground (equivalent to a depth $X_{\mbox{decay}}$), where it creates a shower.}
  \label{fig:neutrinoscheme}
\end{figure}

Due to their nature, these neutrino initiated showers require extra steps in the simulation, such as sampling the position where the neutrino interacts in the atmosphere (see left panel of Fig.~\ref{fig:neutrinoscheme}) and in the case of $\nu_\tau$s also the propagation of the $\tau$ from its creation to its decay, where a shower is created (see right panel of Fig.~\ref{fig:neutrinoscheme}). Since the cross-section of neutrino interactions is very small, even at the highest energies, we assume that the point where the neutrino interaction occurs is equally distributed in atmospheric depth. So, in order to sample the interaction position along the shower axis for the events, we simply divide the atmosphere in slices of equal thickness $\Delta X$ in atmospheric depth, each centered at a different interaction depth $X_{\mbox{int}}$ measured from the top of the atmosphere. This means that we have many more slices at low altitude (high $X_{\mbox{int}}$), due to the higher air density. For each of these slices we perform ZHAireS simulations of showers initiated by CC or NC interactions. From these simulations, instances of the superposition emission model are then created for each of the slices. During the run, for each event the interaction point $X_{\mbox{int}}$ along the shower axis is then sampled by just choosing one of the available atmospheric slices randomly. RDSim then chooses one of the corresponding instances of the emission model at that particular $X_{\mbox{int}}$ slice to simulate the radio emission~\footnote{The thickness $\Delta X$ of the slices is settable. Smaller values of $\Delta X$ will sample the atmosphere more finely, but will require many more ZHAireS simulations to setup an RDSim run. The biggest impact of using a large $\Delta X$ is at the highest altitudes. Due to the lower air density this will lead to large distances (or equivalently large differences in altitude) between the interaction points of the available simulations.}.

In the case of showers initiated by tau-lepton decays (see right panel of Fig.~\ref{fig:neutrinoscheme}), the propagation of the tau is handled in a very simple way, disregarding the $\tau$ energy losses in air. First we sample the depth $X_0$ of the $\nu_\tau$ interaction in the atmosphere. Just as before, this is done by choosing a random atmospheric $\Delta X$ slice. The distance $L_{\mbox{decay}}(E_{\tau})$ between the $\nu_\tau$ interaction and the tau-lepton decay is then sampled by propagating the tau (of energy $E_{\tau}$) in  steps of length $\Delta L$ between the interaction position and the ground. For this we use the probability of tau decay per meter: $dP(E_{\tau})=\frac{m_{\tau}}{E_{\tau}\tau}$, where $\tau=86.93\mu m$ is the decay length. If the $\tau$ decays above ground, this will give us the position along the axis where the tau decays and the shower starts, described by the decay depth $X_{\mbox{decay}}$ measured from the top of the atmosphere. If it does not decay above ground, no shower is created. For speed, the tau propagation simulation is performed externally and prior to the main RDSim run. We record the fraction of taus that do not decay before reaching the ground and thus creates no shower. For those that do decay above ground we create parametrizations of the distributions of $X_{\mbox{decay}}$, which also takes into account the position $X_0$ of the initial interaction of the $\nu_\tau$, where the tau-lepton is produced. These are recorded in text files that are read by RDSim during the main simulation. To create the instances of the superposition emission model to be used in tau decay events, just as before, the atmosphere is divided in slices of equal thickness $\Delta X$ and for each slice ZHAireS simulations are performed. But in this case the products of the tau decay, obtained from Tauola, are injected into ZHAireS instead. During the main run, for each tau decay event we sample the previously obtained probability of the tau decaying above ground. If it does decay above ground the position of the decay is sampled from the previously produced parametrizations. RDSim then chooses one of the corresponding instances of the emission model at the particular slice where the decay occurs to simulate the radio emission. If the tau does not decay before reaching the ground, no shower is created and the event is instantly marked as not triggered. On the left panel of Fig.~\ref{fig:examples} we show an example tau decay event simulated at the AUGER-RD array~\cite{AugerPrime}.

\section{Structure and procedure}
\label{sec:structure}

RDSim was build around speed, but offers comprehensive and precise simulations of the radio emission and its detection by an arbitrary radio array. It was implemented in C++ and is structured around a few key classes, along with tools to help setup a run and analyze the results.
At the end of a run, the information of every event, including event number, arrival direction, core position, Energy, $\sin{\alpha}$, number of triggered stations and the full information about the instance of the emission toymodel used
. In the case of neutrino events, the interaction (or tau decay) height is also saved. For those events that triggered, the full information on all triggered stations is saved, including the electric field components.

The first step to run an RDSim simulation is the creation of multiple instances of the emission model that are relevant for the type of events we want to study. The effect of shower-to-shower fluctuations can be included by just running multiple ZHAireS simulations to create multiple instances of the emission toymodel for the same class of event. This will automatically include the effect of e.g. an \xmax distribution in the simulation. The typical runtime for ZHaireS simulations for this purpose is just $\sim15$ minutes per antenna in a single core, making it possible to create hundreds of emission toymodel instances in just a few hours, if a cluster with $\sim 100$ cores is used.

The main run is controlled by an input file which contains the parameters of the simulation to be run, such as the ranges for the arrival direction, energy and core position, as well as the trigger thresholds. It also lists the emission toymodel instances that are to be included, along with settable ranges for the azimuth angle and energy each one is allowed to be used for\footnote{In general it is safe to allow a toymodel to be rotated to any azimuth angle and to be scaled to energies at least half an order of magnitude around its original energy.}. This main input file also contains the location of the optional files for the antenna beam patterns and, in the case of tau decay events, the parametrizations previously obtained for the tau propagation. During the run, for each event we sample an isotropic arrival direction (with zenith rounded up to the nearest available toymodel), along with an energy and a core position equally distributed in area. RDSim then searches for all toymodels that can be used for these particular values and chooses one of them randomly. The chosen toymodel is rotated and scaled to match the parameters of the event and then used to calculate the electric field at each antenna. The beam pattern is then applied to the calculated fields and the trigger condition is checked for each antenna.

\section{Discussion}
\label{sec:discussion}

\begin{figure}[!htb]
  \begin{center}
    \includegraphics[width=0.352\linewidth]{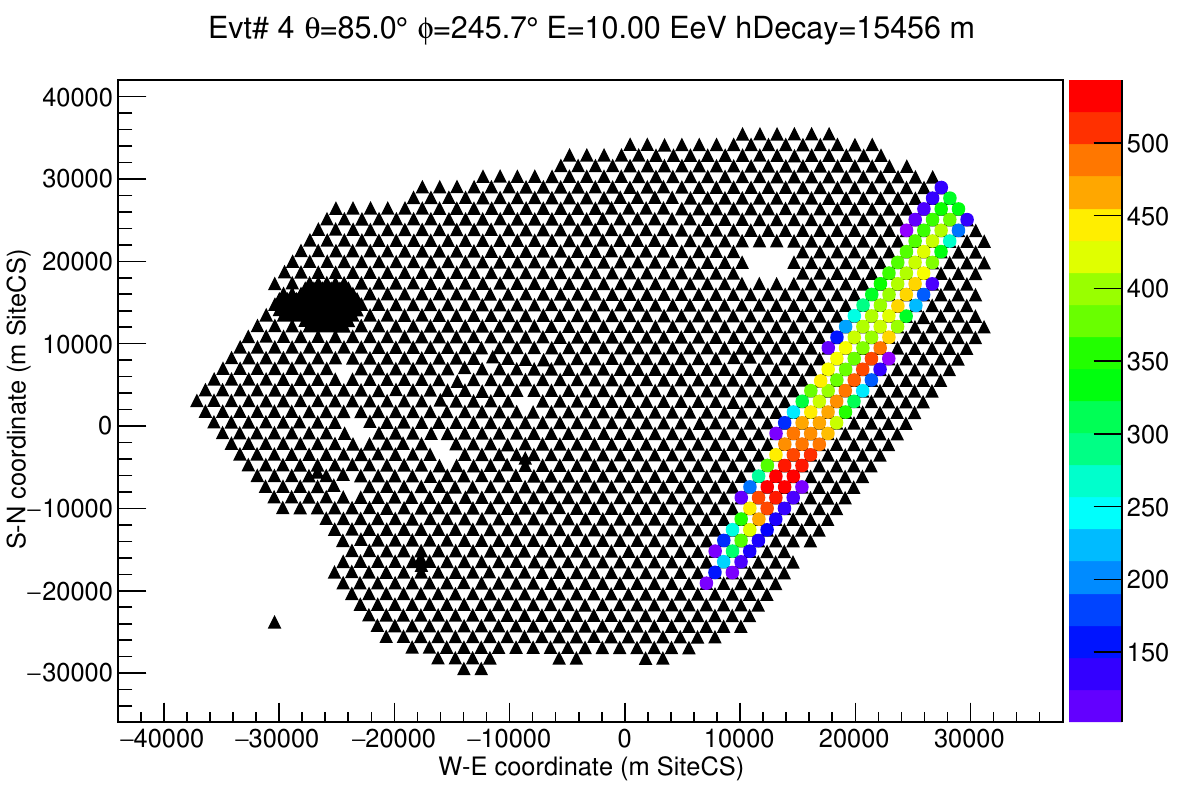}\includegraphics[width=0.352\linewidth]{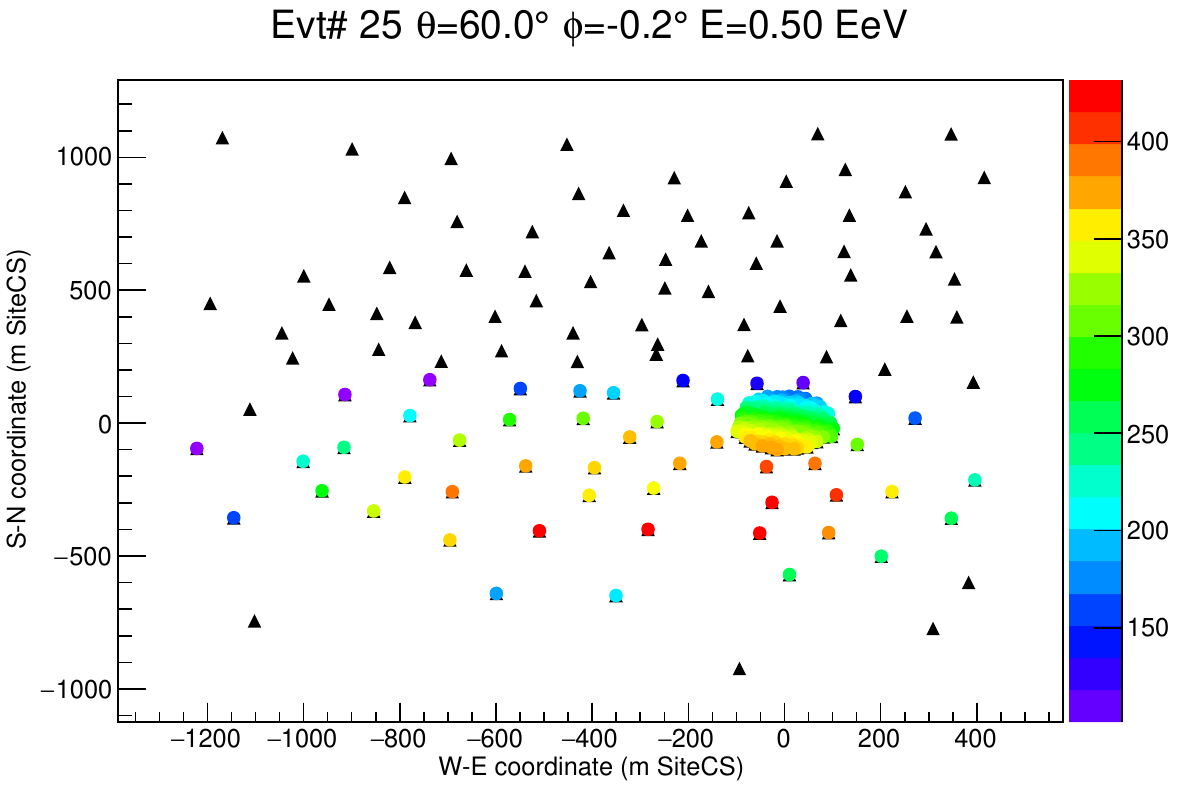}\includegraphics[width=0.28\linewidth]{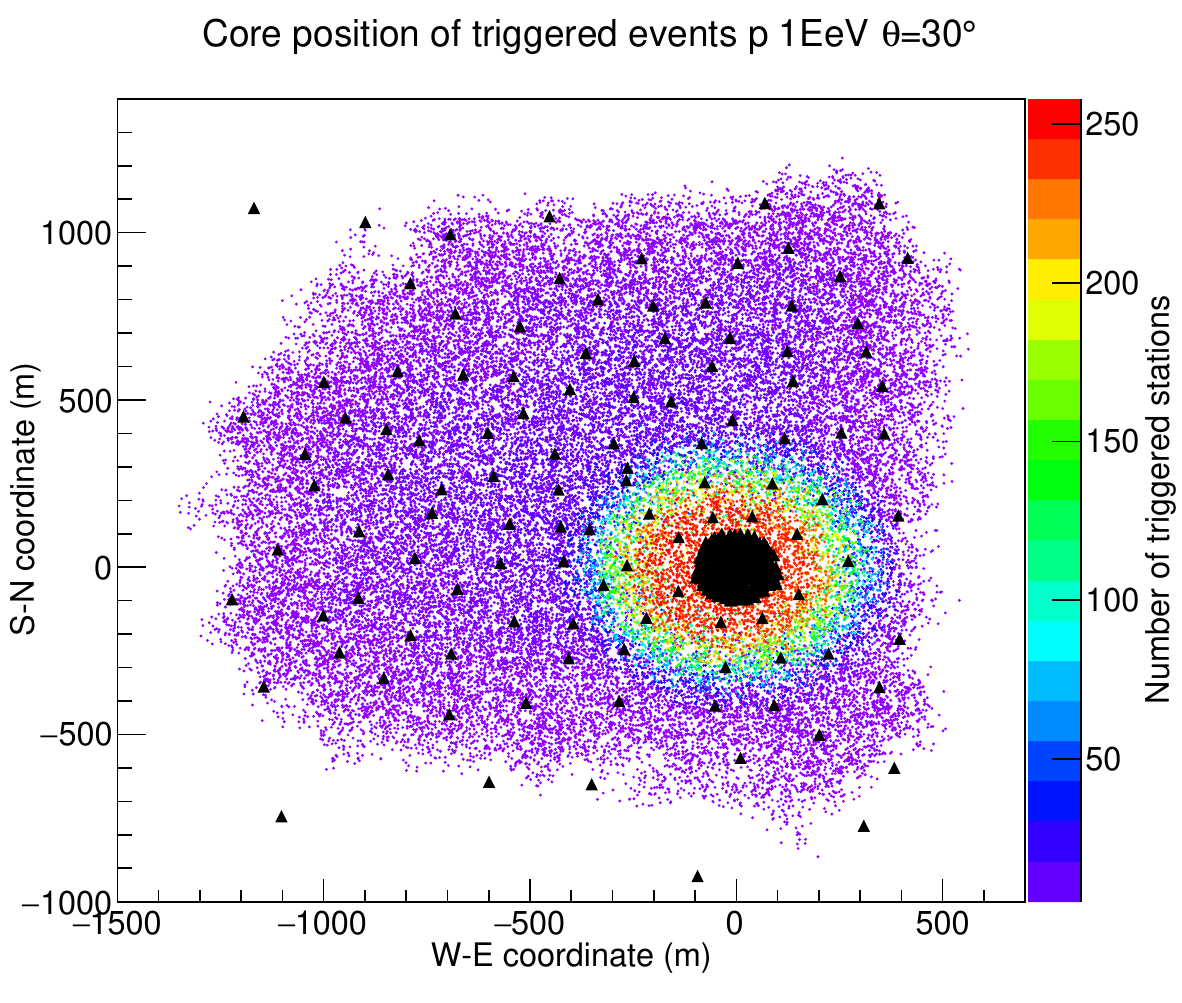}
  \end{center}
  \vspace{-0.6cm}
  \caption{Left: Example event induced by an 85$^\circ$ tau decay occurring $\sim15.5$ km above the AUGER-RD array. Middle: Example of a 60$^\circ$, 0.5 EeV proton event at OVRO-LWA. In both examples the color scale represents the measured electric field in $\mu V/m$. Right: Number of triggered stations as a function of core position for 1 EeV, 30$^\circ$ proton showers at OVRO-LWA. }
  \label{fig:examples}
\end{figure}

RDSim is a very flexible framework for the simulation of radio detection. It can be used to model the response of very different antenna arrays, as illustrated by the left and middle panels of Fig.~\ref{fig:examples}, where we show example events simulated at AUGER-RD (left) and OVRO-LWA (right). Owing to the large statistics made possible by its speed, RDSim can be used to perform detailed studies of the impact of the array characteristics in its detection capabilities. On the right panel of Fig.~\ref{fig:examples} we show a very simple example of such a study, depicting the number of triggered stations as a function of core position for 30$^\circ$ 0.5 EeV proton showers at OVRO-LWA. One can see that for the zenith angle studied, events landing in the NW and South parts of the OVRO-LWA array cannot be detected due to the low antenna density in those regions, while the number of triggered stations rises fast as the shower core approaches the center of the array with its very high antenna density.

Events with a low probability of detection can also be investigated in detail.  This can be very important as many different classes of these low probability events can add up to a sizable contribution to e.g. the aperture of a detector. RDsim can also be very useful to generate a much clearer picture of the contribution of each class of event, making it possible to determine what can and cannot be seen by a detector and to what extent, by estimating the detection probability of each class of event. This is particularly useful if one desires to perform more detailed studies based on full simulations, as it can be used to optimize the phase-space to be fully simulated. It can be used to estimate not only the total number of full simulations needed to thoroughly cover the relevant phase-space, but also the relative number of simulations to be performed for each class of event, based on their detection probability. This is specially important for  $\nu$ ($\tau$ decay) induced showers, as they have a much bigger phase space if compared to regular showers due to their extra relevant variables, such as their highly variable interaction (decay) depth, making the generation of unoptimized full simulation libraries unfeasible.

Recently we started comparing RDSim results to full emission and detector simulations. Preliminary results of these comparisons shows a very good agreement, despite the astronomical decrease in the computing time needed by RDSim. We are also in the process of adding the functionality to simulate mountain events, i.e. events induced by the decay of tau-leptons produced by $\nu_\tau$ interactions in the terrain around the detector. For this we will use topographical maps of the region to calculate, given a core position and an arrival direction for an event, the amount of rock traversed and the distance to the closest rock face. This information will then be used after the main run to convolute the probability of detection of a $\tau$ of energy $E_\tau$, established by RDSim, with the probability of such tau-lepton exiting the mountain as a function of the energy $E_\nu$ of the $\nu_\tau$  that produces it.

\end{document}